\documentclass[3p,12pt]{elsarticle}
\usepackage{tabularx} 
\usepackage{graphicx} 
\usepackage[intlimits]{amsmath}
\usepackage{amssymb} 
\usepackage{soul} 
\usepackage{hyperref} 
\usepackage[version=4]{mhchem}
\usepackage{siunitx}
\usepackage[capitalise]{cleveref}
\DeclareSIUnit{\atm}{atm}
\usepackage{appendix}

\begin{document}

\begin{frontmatter}
\title{Role of enthalpy transport in laminar premixed hydrogen flames at atmospheric and elevated pressures}

\author{T.~L.~Howarth\corref{cor1}}
\author{T.~Lehmann}
\author{M.~Gauding}
\author{H.~Pitsch}
\journal{XXX}

\address{Institute for Combustion Technology, RWTH Aachen University, Templergraben 64, 52056 Aachen, Germany}
\cortext[cor1]{Corresponding author: t.howarth@itv.rwth-aachen.de}

\begin{abstract} 
This work discusses the role of diffusive enthalpy transport in relation to the origin of thermodiffusive instability and the resulting enhanced reactivity. Thermodiffusive effects in premixed hydrogen flames are typically explained and modelled via local equivalence ratio fluctuations. However, it is reiterated here that the imbalance between species and thermal diffusion (differential diffusion), rather than local species-to-species diffusive imbalances (preferential diffusion) is the leading-order effect. Reactant (\ce{H2}), product (\ce{H2O}) and intermediate (\ce{H}) species are demonstrated to all play a role in the transport of enthalpy through an analysis of enthalpy flux divergence terms in unstretched flames. Premixed counterflow flames at various strain rates and pressures are then analysed to demonstrate that enhanced reactivity originates from a combination of enthalpy transport and the broadness of the reaction zone relative to the thickness of the flame. Effects resulting from key pressure fall-off reactions are also discussed to determine the importance of detailed chemistry, and the usage of Zeldovich number. Finally, two-dimensional planar flames are simulated and analysed to demonstrate the role of curvature in addition to strain rate, and the implications of the findings in blends and turbulent flames are discussed. 
\end{abstract}

\end{frontmatter}





\section{Introduction}
Differential diffusion effects in premixed flames have recently gained a great deal of attention as a result of the ongoing transition from carbon-based to hydrogen-based fuels. Due to its small molecular weight, hydrogen diffuses rapidly, and fuel/air mixtures involving hydrogen can be subject to thermodiffusive instabilities, which drastically increase local burning rates depending on the mixture composition, temperature, pressure and turbulence level~\cite{aspden2011turbulence,howarth2022empirical,berger2022intrinsicb,berger2022synergistic,howarth2023thermodiffusively}.

Thermodiffusive effects are typically described through the fluctuation of local mixture fraction/equivalence ratio (e.g.~`richer'/`leaner' burning in positive/negative curvature regions). Modelling efforts have also focused on this interpretation, and several models have been developed based on progress variable and mixture fraction~\cite{regele2013two,berger2022development,yao2024capturing}. However, it was demonstrated by \citet{aspden2017numerical} that in turbulent flames, even when all species Lewis numbers are set equal to that of hydrogen, suppressing local fluctuations in the mixture fraction, the thermodiffusive response of the flame remains unchanged. Here, the distinction is made between preferential diffusion, i.e.~species-to-species diffusive competition (non-equal Lewis numbers), versus differential diffusion, which is the imbalance between species and thermal diffusion across the flame (non-unity Lewis numbers). Although preferential diffusion is quantified through local fluctuations in mixture fraction, differential diffusion is quantified through local fluctuations in enthalpy. Efforts have also been made to formulate flamelet-based models with enthalpy as a parameter \cite{bottler2022flamelet,bottler2023flamelet,bottler2024can}. However, exploration of the fundamental role of enthalpy fluctuations and the transport processes leading to these phenomena, whether laminar or turbulent, has not been considered.

Under the assumption of mixture-averaged diffusivities, the diffusive flux for a species $k$ is given by
\begin{equation}\label{eq:spec_diff}
        \mathbf{F}_{k} = -\frac{\rho D_{k,mix}\overline{W}}{W_{k}}\nabla Y_{k} -\frac{\rho D_{k,mix}Y_{k}}{W_{k}}\nabla \overline{W} - \frac{\rho D_{k,mix}\chi_{k}}{T}\nabla T + \mathbf{F}_{c}.
\end{equation}
Here, $\rho, T, \overline{W}$ and $\mathbf{F}_{c}$ are the density, temperature, mean molecular weight and mass conservation flux respectively and $D_{k,mix}, Y_{k}, W_{k}$ and $\chi_{k}$ are the mixture-averaged diffusivity, mass fractions, molecular weight and thermal diffusion ratio of species $k$ respectively. If the Soret and molecular weight term are neglected, and all species Lewis numbers are assumed to be unity other than a single deficient species $Y_{F}$ with constant Lewis number $Le_{F}$, a transport equation for mixture fraction $Z$ can be derived~\cite{regele2013two},
\begin{equation}
        \frac{\partial \rho Z}{\partial t} + \nabla \cdot (\rho Z\mathbf{u} + \mathbf{F}_{Z}) = 0, \quad \mathbf{F}_{Z} = -\rho D \nabla Z - \rho D \left(\frac{1}{Le_{F}} -1\right)(1-Z)\nabla Y_{F},
\end{equation}
for thermal diffusivity $D$ and velocity field $\mathbf{u}$. Total specific enthalpy $h$ is also a conserved scalar following a similar transport equation, given by
\begin{equation}\label{eq:enth_transport}
    \frac{\partial \rho h}{\partial t} + \nabla \cdot (\rho h\mathbf{u} + \mathbf{Q}) = 0, \quad \mathbf{Q} = -\lambda \nabla T + \sum_{k}h_{k}\mathbf{F}_{k}
\end{equation}
for thermal conductivity $\lambda$, and species enthalpy $h_{k}$. If the Soret and molecular weight contributions in \cref{eq:spec_diff} are neglected again, then combining \cref{eq:enth_transport} along with the relations
\begin{equation}
    h = \sum_{k}h_{k}Y_{k}, \quad \nabla h_k = c_{p,k}\nabla T
\end{equation}
with species specific heat capacities $c_{p,k}$, gives
\begin{equation}\label{eq:enth_source}
    \mathbf{Q} = -\frac{\lambda}{c_{p}}\nabla h - \sum_{k}h_{k}\rho D_{k}\left( \frac{1}{Le_{k}} - 1\right)\nabla Y_{k}, \quad Le_{k} = \frac{\lambda}{\rho c_{p}D_{k}} 
\end{equation}
where the notation has been simplified to $c_{p} = c_{p,mix}$, $D_{k} = D_{k,mix}\overline{W}/W_{k}$. Unlike the mixture fraction equation derivation, the Lewis numbers for the species have not been assumed constant for each species, and may vary in space.

This work focuses on the analysis of the various flux divergence terms arising from the enthalpy transport equation, i.e.~the decomposition of $-\nabla \cdot \mathbf{Q}$ in \cref{eq:enth_source}. After discussing the solver and direct numerical simulation (DNS) databases, the flux divergence terms are first analysed in unstretched one-dimensional premixed flames at low and high pressure, as thermodiffusive instability is known to be sensitive to the pressure of the system~\cite{howarth2022empirical,berger2022intrinsica,berger2022intrinsicb,rieth2023effect}. Then, enthalpy flux divergence terms are analysed in one-dimensional counterflow premixed flames, again at low and high pressure to explain the effect of strain on the diffusive fluxes and resulting enhanced reactivity. Based on the analysis of the 1D flamelets, the role of detailed chemistry is shortly discussed, before finally analysing DNS at low and high pressure to understand how the findings from 1D translate to 2D and the role of curvature.

\section{Direct numerical simulation solver and database}
Direct numerical simulations of two-dimensional, lean premixed hydrogen flames employing detailed chemistry were performed using PeleLMeX~\cite{PeleLMeX_JOSS}. PeleLMeX solves the reactive Navier-Stokes equations in the low-Mach number limit. The discretisation of these equations couples a multi-implicit spectral deferred correction approach~\cite{nonaka2018conservative} with a density-weighted approximate projection method~\cite{almgren1998conservative}, which incorporates the equation of state and constant thermodynamic pressure through a divergence constraint on the velocity field~\cite{day2000numerical}. Transport properties, thermodynamic relations and chemical rates were taken from a comprehensive mechanism suitable for high-pressure hydrogen flames~\cite{burke2012comprehensive}.

Two cases were considered, both at $\phi = 0.4, T_{u}=\SI{298}{\kelvin}$, with a low-pressure (LP) case at $p=\SI{1}{\atm}$ and a high-pressure (HP) case at $p=\SI{10}{\atm}$. Both cases were set with domain sizes of $200l_{L} \times 400 l_{L}$, where $l_{L}$ is the unstretched laminar flame thickness from the 1D simulation. A large domain size was chosen to avoid confinement effects~\cite{berger2019characteristic} and obtain good statistics. To obtain a resolution of approximately 16 cells across the freely propagating flame thickness~\cite{howarth2022empirical}, a base resolution of $768\times 1536$ with three levels of adaptive mesh refinement was used in the LP case. Since the flame will experience more instability and hence be thinner for the HP case, a higher base resolution of $2048\times 4096$ was used, again with three levels of refinement. Lateral $x$-direction boundaries are set as periodic, with inflow-outflow boundaries set on the $y$-boundaries respectively. A profile was initialised using a one-dimensional unstretched profile from Cantera~\cite{cantera} and a superposition of harmonic functions was applied to the flame at initialisation. This allows instabilities to grow and develop quickly before statistics are taken.

\section{Flamelet behaviour}
In this section, the behaviour of the flux divergence terms in the enthalpy transport equation (\cref{eq:enth_transport}) are analysed in premixed one-dimensional unstretched and counterflow flamelets solved using Cantera. Here, the terms are separated with the following notation
\begin{equation}\label{eq:enth_diff_fluxes}
    \mathbf{Q}_{h} = -\frac{\lambda}{c_{p}}\nabla h, \quad \mathbf{Q}_{k} = -h_{k}\rho D_{k}\left( \frac{1}{Le_{k}} - 1\right)\nabla Y_{k}.
\end{equation}

\subsection{Unstretched flamelets}
\begin{figure}[h!]
    \centering
    \includegraphics[width=0.98\linewidth]{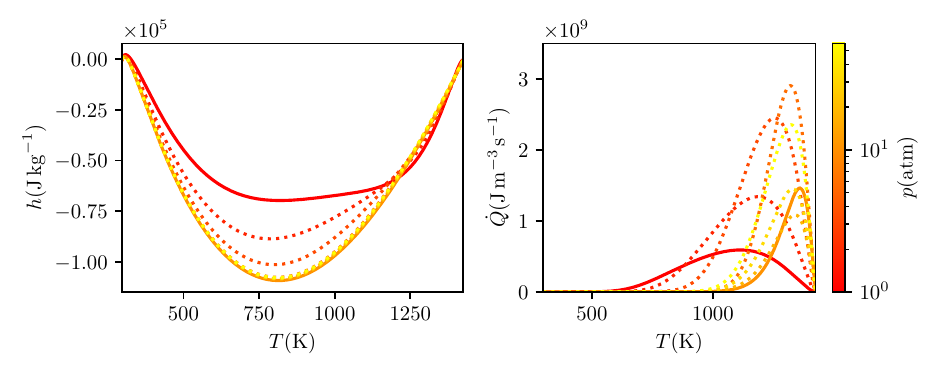}
    \vspace{-4mm}
    \caption{Enthalpy $h$ (left) and heat release rate $\dot{Q}$ (right) in temperature space in an unstretched hydrogen/air flame at $\phi = 0.4, T_{u} = \SI{298}{\kelvin}$ with increasing pressure. The LP and HP cases are denoted by solid lines.}
    \label{fig:enth_hrr}
\end{figure}

\cref{fig:enth_hrr} shows the enthalpy and heat release profiles with increasing pressure as a function of temperature in an unstretched premixed flame at $\phi = 0.4, T_{u} = \SI{298}{\kelvin}$. With increasing pressure, the enthalpy profile approaches a limiting state, whereas the heat release profile initially gets thinner in temperature space with increasing pressure before broadening for $p > \SI{10}{\atm}$. In this section, the analysis is restricted to profiles of the LP and HP cases (denoted by solid lines); a discussion on the general effect of pressure is presented later.

\begin{figure}[h!]
    \centering
    \includegraphics[width=0.98\linewidth]{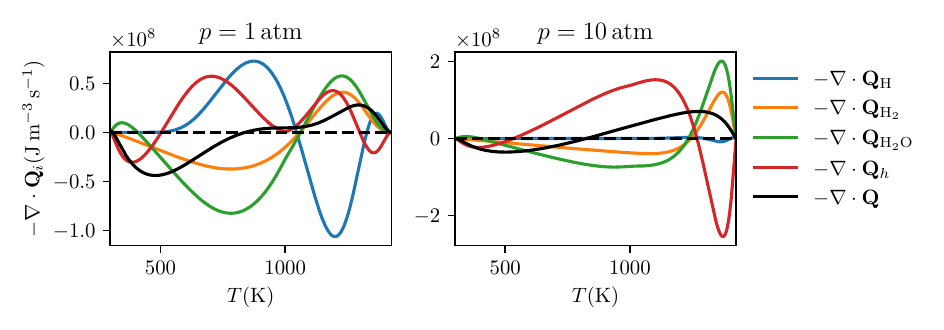}
    \vspace{-4mm}
    \caption{Enthalpy flux divergence terms $-\nabla \cdot \mathbf{Q}_{h}$ and $-\nabla \cdot \mathbf{Q}_{k}, k=\ce{H},\ce{H2},\ce{H2O}$ in temperature space for the LP (left) and HP (right) cases. The dashed line denotes 0.}
    \label{fig:enth_source}
\end{figure}

\cref{fig:enth_source} shows $-\nabla \cdot \mathbf{Q}_{h}$ and $-\nabla \cdot \mathbf{Q}_{k}$ for $k = \ce{H},\ce{H2},\ce{H2O}$ for the LP and HP cases. Since the other species have a near-unity Lewis number, their contributions are relatively small; \cref{fig:enth_source_full} in the supplementary material presents a similar plot but with all species included. $Y_{\ce{H2}}$ has a consistently negative gradient with $Le_{\ce{H2}} \approx 0.3$ and $h_{\ce{H2}} \geq 0$, meaning enthalpy is transported from low to high temperatures. $Y_{\ce{H2O}}$ has a consistently positive gradient, with $Le_{\ce{H2O}} \approx 0.8$ for $T > 400\si{\kelvin}$ and $h_{\ce{H2O}} < 0$, resulting in a transport of negative enthalpy from high to low temperatures, giving a similar effect to \ce{H2}. $Y_{\ce{H}}$ has a sharp central profile in the reaction zone with $Le_{\ce{H}} \approx 0.17$, resulting in enthalpy transport out of the reaction zone into both the low and high temperatures. In particular, \ce{H} counteracts some of the enthalpy transported by \ce{H2} and \ce{H2O}. The net effect of all three species and natural gradient diffusion of enthalpy is the accumulation of enthalpy in the high-temperature region of the flame. The heat release profile for the LP case is broad in temperature space ($\dot{Q}(T) > 0$ for $\SI{600}{\kelvin} < T < T_{ad}$, see \cref{fig:enth_hrr}), and enthalpy is transferred from regions lower than $\SI{900}{\kelvin}$ to regions above this temperature. Therefore, large portions of enthalpy are being redistributed within the reaction zone rather than being transferred from the preheat region.


At high pressure, the structure of the flame is different. A lower overall enthalpy is achieved in the centre of the flame, and the reaction zone is much thinner in temperature space ($\dot{Q}(T) > 0$ for $\SI{1000}{\kelvin} < T < T_{ad}$). At high-pressure conditions, where \ce{H} radicals are consumed through the \ce{H + O2 ( + M) <=> HO2 ( + M)} reaction, the counteracting effect of the \ce{H} radical to the transport of enthalpy into the reaction zone via \ce{H2} and \ce{H2O} is suppressed. A reduction in the \ce{H} radical has resulted in essentially no diffusive enthalpy transport out of the high-temperature regions (see the blue line in the right plot of \cref{fig:enth_source}). The crossover temperature for the transfer of enthalpy from low temperatures to high temperatures in the HP case is similar to the LP case ($T(-\nabla \cdot \mathbf{Q} = 0) \approx \SI{900}{\kelvin})$, but the reaction zone is only situated at temperatures higher than this, and so enthalpy is only being transferred into the reaction zone from the preheat region. 

\subsection{Premixed counterflow flamelets}
\begin{figure}[h]
    \centering
    \includegraphics[width=0.95\linewidth]{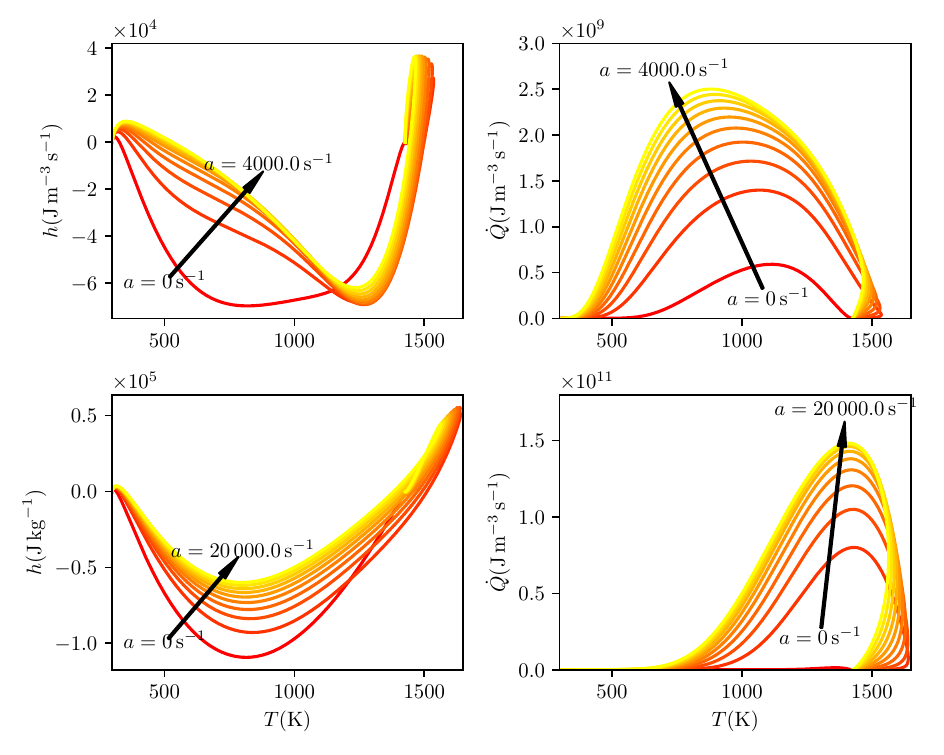}
    \vspace{-4mm}
    \caption{Enthalpy (left) and heat release rate (right) in temperature space in the counterflow LP (top) and HP (bottom) case with increasing strain rate $a$. Note the much higher strain rate applied to the HP case ($a = \SI{20000}{\per\second}$) compared to the LP case ($a = \SI{4000}{\per\second}$).}
    \label{fig:enth_hrr_strain}
\end{figure}

\begin{figure}[h]
    \centering
    \includegraphics[width=0.99\linewidth]{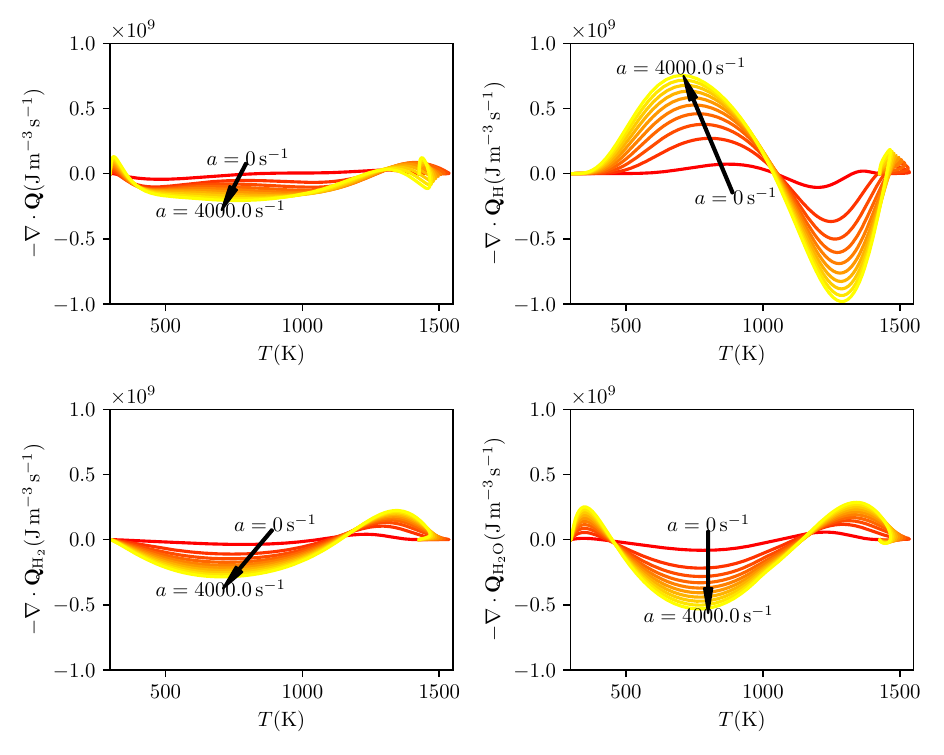}
    \vspace{-4mm}
    \caption{Enthalpy flux divergence terms $-\nabla \cdot \mathbf{Q}$ and key species flux divergence terms $-\nabla \cdot \mathbf{Q}_{\ce{H}},-\nabla \cdot \mathbf{Q}_{\ce{H2}},-\nabla \cdot \mathbf{Q}_{\ce{H2O}}$ in temperature space for the LP case with increasing strain rate $a$.}
    \label{fig:enthalpy_strain_decomp}
\end{figure}

In a reactants-to-products premixed counterflow flamelet, a well-defined strain rate can be applied to the flame. By applying strain, diffusive fluxes are enhanced due to the stretching of the control volume within the flame \cite{law2000structure}. Therefore, the flux divergence terms in the enthalpy transport equation \cref{eq:enth_diff_fluxes} can be expected to be enhanced by strain. \cref{fig:enth_hrr_strain} shows the enthalpy and heat release profile in temperature space for the LP and HP cases with increasing strain values. As the strain rate increases, the enthalpy profile shifts to a new limiting state. The heat release also distinctly increases due to the increased local enthalpy. To understand the mechanism for this shift, the enthalpy flux divergence term $-\nabla \cdot \mathbf{Q}$, as well as the key species terms $-\nabla \cdot \mathbf{Q}_{k}, k= \ce{H},\ce{H2},\ce{H2O}$ are shown in \cref{fig:enthalpy_strain_decomp} for the LP case. As strain rate increases, the overall enthalpy flux divergence term $-\nabla \cdot \mathbf{Q}$ increases at the extreme ends of the temperature range ($T < \SI{400}{\kelvin}, T > \SI{1400}{\kelvin}$), while it decreases through the flame. As with the unstretched flame, if the heat release profile in \cref{fig:enth_hrr_strain} and the flux divergence terms in \cref{fig:enthalpy_strain_decomp} are compared, it can be seen that at the highest strain rate ($a = \SI{4000}{\per\second}$), enthalpy is mostly being lost from the reaction zone (which is now the vast majority of the flame) to the reactants and products and hence reactivity cannot be expected to increase further with increasing strain. As can also be seen in \cref{fig:enthalpy_strain_decomp}, the effects observed in the unstretched flame for each species, i.e.~transport of enthalpy to low temperatures through the \ce{H} radical, and high temperatures by \ce{H2} and \ce{H2O}, are enhanced by strain. At the highest strain rates, enthalpy is sufficiently lost from both the high temperature through \ce{H}, and at the lower temperatures through \ce{H2} and \ce{H2O} that reactivity cannot increase any further.

\begin{figure}[h]
    \centering
    \includegraphics[width=0.98\linewidth]{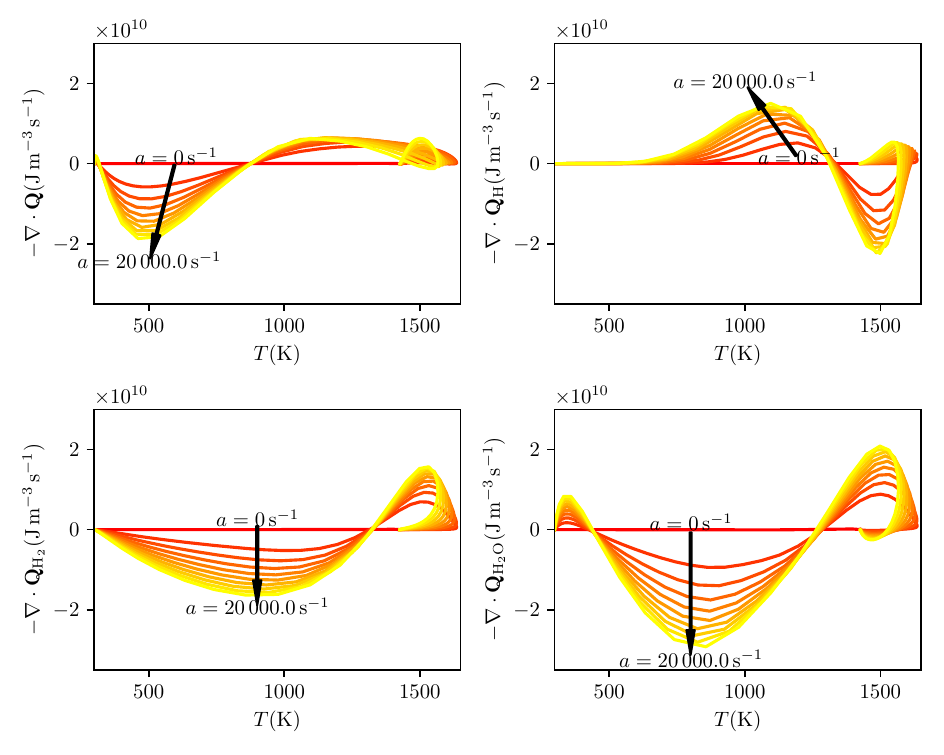}
    \vspace{-4mm}
    \caption{Enthalpy flux divergence term $-\nabla \cdot \mathbf{Q}$ and key species flux divergence terms $-\nabla \cdot \mathbf{Q}_{\ce{H}},-\nabla \cdot \mathbf{Q}_{\ce{H2}},-\nabla \cdot \mathbf{Q}_{\ce{H2O}}$ in temperature space for the HP case with increasing strain rate $a$.}
    \label{fig:enthalpy_strain_highP_decomp}
\end{figure}

In the HP case, also shown in \cref{fig:enth_hrr_strain}, a far higher strain rate needs to be applied to the flame before a limiting profile is seen, suggesting that the flame is much more resilient to higher strain rates. This cannot be simply explained by the difference in timescales, as the flame time for the HP case is approximately 10 times larger than the LP case, resulting in normalised strain rates that are 50 times larger. Again, by comparing the heat release profile in \cref{fig:enth_hrr_strain} and the different flux divergence terms shown in \cref{fig:enthalpy_strain_highP_decomp}, it can be seen that even at the highest strain rate, there are still considerable enthalpy gains in the high-temperature ($T > 1000\si{\kelvin}$) regions. Since the reaction zone is not as broad as the low-pressure case, and transport due to the \ce{H} radical is still not very large, reactivity can continue to increase. 

\subsection{Single-step vs. detailed chemistry effects}
Two effects contribute to the degree of enhanced reactivity. Firstly, the transport of enthalpy from low to high temperatures is key, and the extent of the redistribution of enthalpy within the reaction zone determines the potential for enhancement. At low pressures, the reaction zone is spread over a large temperature range (\cref{fig:enth_hrr}), and the net effect of the enthalpy transport is mostly redistribution within the reaction zone. However, at high pressures, the combined thin reaction zone and consumption of \ce{H} means enthalpy is only accumulated at high temperatures, and the flame reactivity can continue to be enhanced by strain. Both reaction zone thinning and consumption of the \ce{H} radical are due to the pressure sensitivity of the \ce{H + O2 ( + M) <=> HO2 ( + M)} reaction \cite{law2006propagation}, and so the effects are coupled. However, at even higher pressures ($p > \SI{10}{\atm}$, at $\phi = 0.4, T_{u}=\SI{298}{\kelvin}$), the effects begin to decouple as the reaction zone starts to broaden again due to the activation of the chain-branching reaction \ce{H2O2 ( + M) <=> 2OH ( + M)} which increases the reactivity, allowing the reaction zone to extend to lower temperatures. However, the enthalpy profile is essentially unchanged; these effects can be seen in \cref{fig:enth_hrr}. As the flame transitions to this high-pressure regime, instability would be expected to decrease as the reaction zone shifts to lower temperatures where enthalpy is being redistributed to higher temperatures, however, the \ce{H} radical is still consumed, preventing enthalpy transport from high to low temperatures. Therefore, at high pressures enthalpy transport is only driven by diffusion of reactant (\ce{H2}) and product (\ce{H2O}) species and detailed chemistry effects are unimportant; only the broadness of the reaction zone relative to the flame thickness (i.e.~Zeldovich number effect) is important. This explains the linear relationship between thermodiffusive response and Zeldovich number found by \citet{howarth2022empirical} at high pressure, whereas the correlation at low pressures was found to be non-linear.

\section{Multi-dimensional behaviour}
\begin{figure}[h!]
    \centering
    \includegraphics[width=0.98\linewidth]{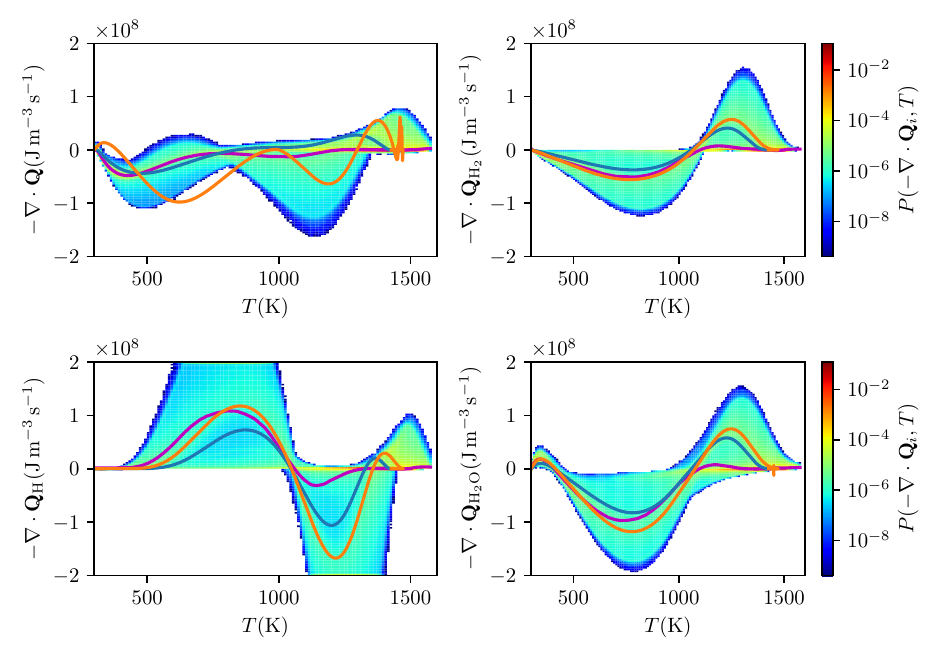}
    \vspace{-4mm}
    \caption{JPDFs of enthalpy flux divergence term $-\nabla \cdot \mathbf{Q}$ and key species flux divergence terms $-\nabla \cdot \mathbf{Q}_{k},k=\ce{H},\ce{H2},\ce{H2O}$ with temperature for the LP case, along with corresponding conditional means (magenta), unstretched flamelet (blue) and weakly strained flamelet (orange, $a= \SI{70}{\per\second}$).}
    \label{fig:enth_source_jpdf}
\end{figure}

\begin{figure}[h!]
    \centering
    \includegraphics[width=0.98\linewidth]{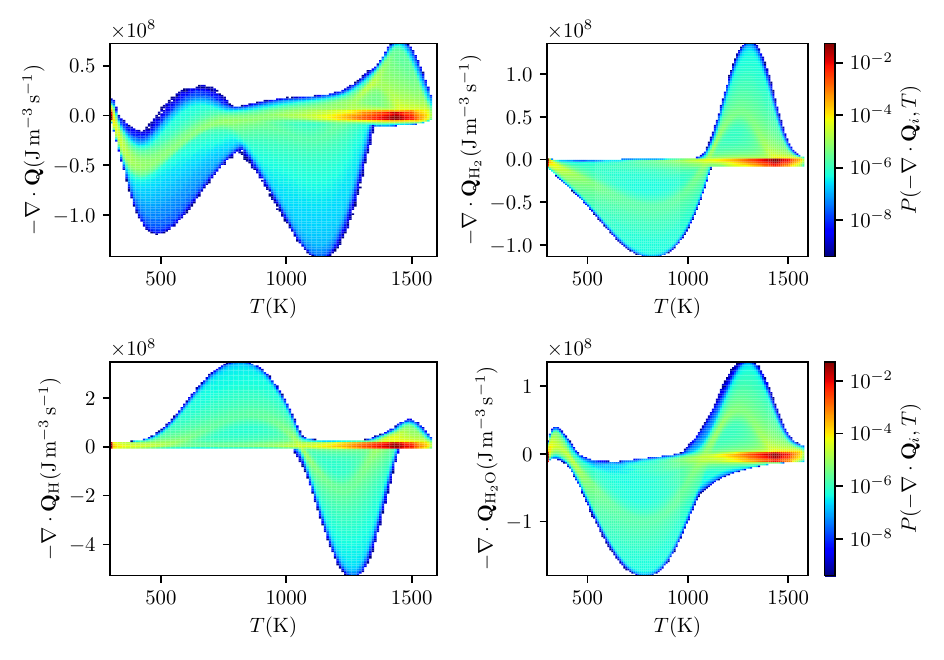}
    \caption{Extended version of \cref{fig:enth_source_jpdf} showing the full range of JPDF of the enthalpy flux divergence terms for the LP case.}
    \label{fig:enth_source_jpdf_full}
\end{figure}

In more than one dimension, perturbations to the flame surface will enhance diffusive fluxes through both curvature and strain rate, and the flame will develop characteristic structures typical of thermodiffusive instability \cite{berger2019characteristic}. 
\begin{figure}[h]
    \centering
    \includegraphics[width=0.98\linewidth]{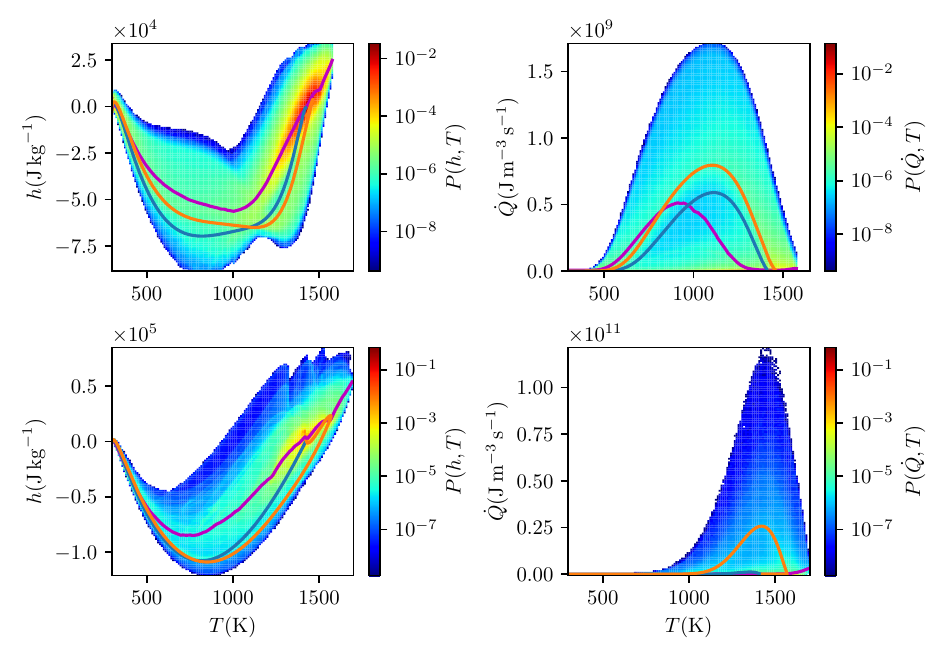}
    \vspace{-4mm}
    \caption{JPDFs of enthalpy (left) and heat release rate (right) with temperature for the LP and HP cases, along with corresponding conditional means (magenta), unstretched flamelet (blue) and weakly strained flamelet (orange, $a_{\rm LP} = \SI{70}{\per\second}, a_{\rm HP} = \SI{200}{\per\second}$).}
    \label{fig:enth_hrr_jpdf}
\end{figure}
\cref{fig:enth_hrr_jpdf} shows the joint probability distribution functions (JPDFs) of enthalpy and temperature and of heat release and temperature along with corresponding conditional means (magenta), unstretched laminar flame profile (blue) and a weakly strained flame (orange, $a_{\rm LP}=\SI{70}{\per\second},a_{\rm HP}=\SI{200}{\per\second} $) for the LP and HP cases in a 2D planar laminar flame. In the LP case, the flame covers a wide range of enthalpy for a given value of temperature, and the conditional mean aligns with neither the unstretched nor the weakly strained flamelet. To explain these different states, \cref{fig:enth_source_jpdf} shows the various enthalpy divergence flux terms for the LP case. The conditional mean for the overall enthalpy flux divergence (top left panel of \cref{fig:enth_source_jpdf}) has only shifted a small amount from the unstretched flamelet. However, significant fluctuations can be seen through the flame. In particular, there are noticeable extra regions of enthalpy loss at $\SI{300}{\kelvin} < T < \SI{700}{\kelvin}$ and $\SI{900}{\kelvin} < T < \SI{1300}{\kelvin}$ and enthalpy gain for $T > \SI{1300}{\kelvin}$. A similar qualitative trend can be seen in the weakly strained flamelet, but the profiles do not line up. By examining the species terms in \cref{fig:enth_source_jpdf}, these fluctuations can be explained similarly to the counterflow flame case. \ce{H2} and \ce{H2O} transport enthalpy from the low-temperature regions into the high-temperature regions, whereas the \ce{H} radical has the opposite effect. A figure showing the full range of the JPDF is given as supplementary material (\cref{fig:enth_source_jpdf_full}).

\begin{figure}[h!]
    \centering
    \includegraphics[width=0.98\linewidth]{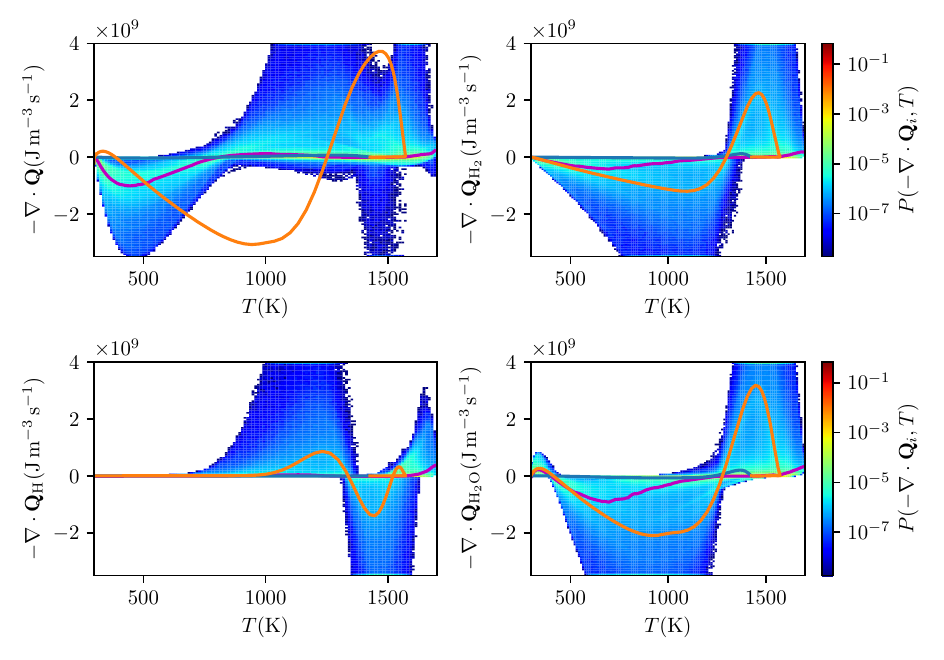}
    \vspace{-4mm}
    \caption{JPDFs of enthalpy flux divergence term $-\nabla \cdot \mathbf{Q}$ and key species flux divergence terms $-\nabla \cdot \mathbf{Q}_{k},k=\ce{H},\ce{H2},\ce{H2O}$ with temperature for the HP case, along with corresponding conditional means (magenta), unstretched flamelet (blue) and weakly strained flamelet (orange, $a = \SI{200}{\per\second}$).}
    \label{fig:enth_source_highp_jpdf}
\end{figure}

\begin{figure}[h!]
    \centering
    \includegraphics[width=0.98\linewidth]{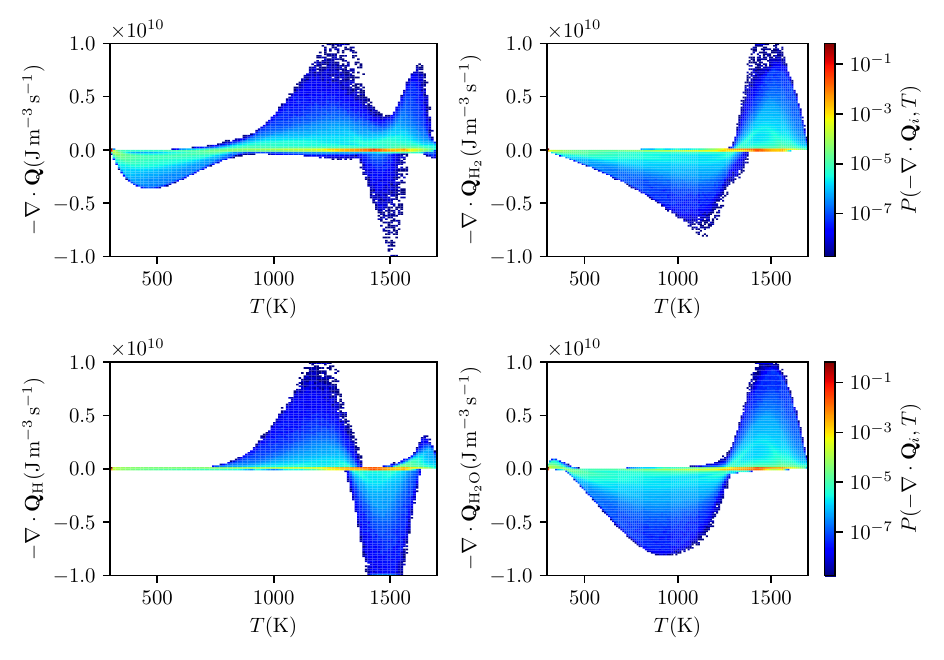}
    \caption{Extended version of \cref{fig:enth_source_highp_jpdf} showing the full range of JPDF of the enthalpy flux divergence terms for the HP case.}
    \label{fig:enth_source_jpdf_highp_full}
\end{figure}

The distribution is different for the HP case. As can be seen in \cref{fig:enth_hrr_jpdf}, for a given value of temperature, the enthalpy distribution of the flame is relatively unskewed and the standard deviation is essentially constant beyond $T = \SI{750}{\kelvin}$; this contrasts with the LP case, where considerable skewness and variation in standard deviation can be seen. This behaviour can be explained by comparing \cref{fig:enth_source_jpdf} and \cref{fig:enth_source_highp_jpdf}. Although a greater range of overall enthalpy flux divergence terms values is seen in the HP case, the probabilities of extreme values are low. In the LP case, there is a relatively even distribution across the flux divergence terms, resulting in a broad distribution of enthalpy values through the flame. Again, a figure (\cref{fig:enth_source_jpdf_highp_full}) is provided as supplementary material showing the full range of the JPDFs for the various flux divergence terms.  


To determine the effect of flame topology and isolate the effects of curvature, an isosurface at $T=\SI{1000}{\kelvin}$ is constructed, and flame normals are extracted following gradients of temperature away from this surface, following the method from \citet{day2009turbulence}. JPDFs are then constructed by organising the flame normals into three categories in the LP case: positively curved regions ($\kappa > \SI{50}{\per\metre}$), negatively curved regions ($\kappa < -\SI{50}{\per\metre}$) and flat regions ($-\SI{50}{\per\metre} < \kappa < \SI{50}{\per\metre}$), where the curvature $\kappa$ has been calculated at the $T=\SI{1000}{\kelvin}$ isosurface. 

\begin{figure}
    \centering
    \includegraphics[width=0.98\linewidth]{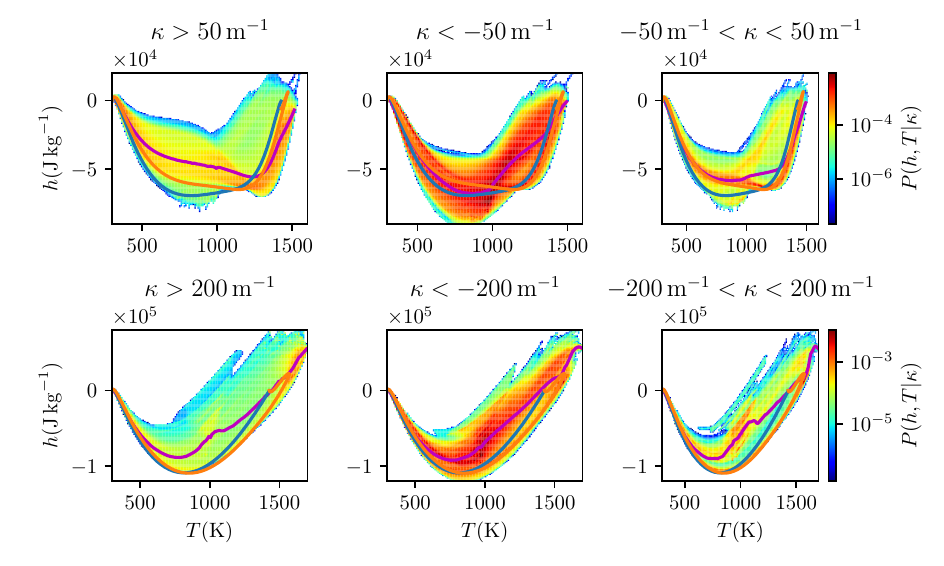}
    \vspace{-4mm}
    \caption{JPDFs of enthalpy with temperature in positively curved (left), negatively curved (centre) and flat regions (right) in the LP (top row) and HP (bottom row) case, along with corresponding conditional means (magenta), unstretched flamelet (blue) and weakly strained flamelet (orange, $a_{\rm LP} = \SI{70}{\per\second}, a_{\rm HP} = \SI{200}{\per\second}$).}
    \label{fig:k_lowp}
\end{figure}

\cref{fig:k_lowp} shows the JPDF of enthalpy and temperature in each category for the LP and HP cases. In the LP case, regardless of the curvature, there is still a significant variation of the profile. However, certain trends can be seen. Specifically, in the flat regions of the flame, a weakly strained ($a=\SI{70}{\per\second}$) flamelet can match the modal profile of enthalpy, although not the conditional mean. In the positive curvature regions, there are heightened enthalpy values due to much higher concentrations of \ce{H} transporting enthalpy to the lower temperatures. In the negative curvature regions, the distribution of enthalpy is much broader, with a very large standard deviation of enthalpy for a given temperature value.

In the HP case, different bounds are used to classify the flame segments: $\kappa  > 200\si{\per\metre}$, $-200 \si{\per\metre} < \kappa < 200 m^{-1}$, $\kappa < -200 \si{\per\metre}$ for the positive curvature, flat and negative curvature regions respectively. Again, in the flat regions, the strained profile (this time $a=\SI{200}{\per\second}$) can recover the modal profile but not the conditional mean due to the significantly higher enthalpy regions in the centre of the flame. Unlike in the LP case, profiles in each region are not that distinct, with conditional means appearing very similar in all regions. This concurs with the idea that at high pressures, the enthalpy profile is relatively insensitive to the local chemistry, unlike at lower pressures.

\section{Discussion and conclusions}
Enthalpy transport behaviour in laminar premixed hydrogen flames has been explored through both unstretched and strained counterflow one-dimensional premixed flames and a database of two-dimensional freely propagating flames. In the unstretched flamelet, \ce{H2}, \ce{H} and \ce{H2O} are the primary drivers for enthalpy transport inside the flame. \ce{H2} and \ce{H2O} act to effectively transport enthalpy into the high-temperature regions, while \ce{H} transports enthalpy out of high-temperature regions into the products and low-temperature regions. At low pressures, this results in a redistribution of enthalpy in the reaction zone, as the heat release profile is distributed over a large range of temperatures. At high pressures, the combination of the suppression of enthalpy transport from \ce{H} and a thin reaction zone situated only at high temperatures results in the accumulation of enthalpy in the reaction zone. The behaviour of the reaction zone can be incorporated into models and theories employing single-step chemistry through the Zeldovich number. However, the involvement of the hydrogen radical in enthalpy transport cannot, meaning unless combustion takes place in regimes where hydrogen radicals are sufficiently suppressed (e.g.~extreme pressures), detailed chemistry effects cannot be ignored.


Once strain is applied to the flame in a reactants-to-products counterflow configuration, diffusive fluxes are enhanced. Since the enthalpy accumulates at high temperatures, reactivity increases and the \ce{H} radical pool increases. Hence, with increasing strain at low pressures, sufficient \ce{H} radical is produced to both broaden the reaction zone and transport enthalpy out of the high-temperature regions. This combination drags the reaction zone into the enthalpy loss region at lower temperatures and reduces the enthalpy accumulation at high temperatures, and beyond a particular strain rate, overall reactivity begins to decrease. At high pressure, the combination of a thin reaction zone and rapid consumption of \ce{H} radicals means that extremely high strain rates are needed before the aforementioned effects take place and flame reactivity starts to decrease. 

Similar behaviour is found in the two-dimensional flame. Enhanced diffusive fluxes resulting from the curvature and strain rate in the LP case result in larger enthalpy gains in the high-temperature regions, driving thermodiffusive instability. A combination of a thicker reaction zone and larger \ce{H} concentrations prevents the flame reactivity from continuing to increase. In the HP case, the thinner reaction zone and reduced \ce{H} concentration enhance reactivity, and a much smaller deviation of the enthalpy distribution is seen. 

These findings have implications for hydrogen fuel blends and turbulent flames, which will be the focus of future work. Firstly, enthalpy fluxes are shown to be sensitive to the \ce{H} radical, which means blend chemistry will play an important role and lead to different correlations between thermodiffusive response and Zeldovich number, as seen by \citet{howarth2024thermal} for \ce{CH4}/\ce{H2} mixtures and \citet{lehmann2025comprehensive} for \ce{NH3}/\ce{H2} mixtures. In turbulent flames, the resilience of high-pressure flames to strain suggests that synergistic interactions will remain up to high turbulence levels and that transitioning to distributed burning will be difficult under these conditions.

\section{Acknowledgements}
TLH acknowledges support from DFG (IRTG 2983 Hy-Potential: Hydrogen - Fundamentals of Production, Storage \& Transport, Applications, and Economy). TL, MG and HP acknowledge support from the European Research Council (ERC) Advanced Grant (HYDROGENATE, ID: 101054894). This project used computing time provided by the Gauss Centre for Supercomputing e.V. (www.gauss-centre.eu) on JUWELS at J\"ulich Supercomputing Centre (JSC) (ID: h2ex). Additionally, computing time was provided by NHR Center NHR4CES at RWTH Aachen University (ID: p0020340). This is funded by the Federal Ministry of Education and Research, and the state governments participating on the basis of the resolutions of the GWK for national high-performance computing at universities (www.nhr-verein.de/unsere-partner).

\section{References}

{\bibliographystyle{mcs13}
\setlength{\bibsep}{0.5mm}
\def\section*#1{}
\bibliography{library}}

\begin{thebibliography}{26}
\newcommand{\enquote}[1]{``#1''}
\providecommand{\natexlab}[1]{#1}

\bibitem[Aspden et~al.(2011)]{aspden2011turbulence}
Aspden, A.J., Day, M.S., Bell, J.B., \enquote{Turbulence--flame interactions in lean premixed hydrogen: transition to the distributed burning regime}, \emph{J. Fluid Mech} 680: 287--320 (2011).

\bibitem[Howarth and Aspden(2022)]{howarth2022empirical}
Howarth, T.L., Aspden, A.J., \enquote{An empirical characteristic scaling model for freely-propagating lean premixed hydrogen flames}, \emph{Combust. Flame} 237: 111805 (2022).

\bibitem[Berger et~al.(2022{\natexlab{a}})]{berger2022intrinsicb}
Berger, L., Attili, A., Pitsch, H., \enquote{{Intrinsic instabilities in premixed hydrogen flames: parametric variation of pressure, equivalence ratio, and temperature. Part 2--Non-linear regime and flame speed enhancement}}, \emph{Combust. Flame} 240: 111936 (2022{\natexlab{a}}).

\bibitem[Berger et~al.(2022{\natexlab{b}})]{berger2022synergistic}
Berger, L., Attili, A., Pitsch, H., \enquote{Synergistic interactions of thermodiffusive instabilities and turbulence in lean hydrogen flames}, \emph{Combust. Flame} 244: 112254 (2022{\natexlab{b}}).

\bibitem[Howarth et~al.(2023)]{howarth2023thermodiffusively}
Howarth, T.L., Hunt, E.F., Aspden, A.J., \enquote{Thermodiffusively-unstable lean premixed hydrogen flames: Phenomenology, empirical modelling, and thermal leading points}, \emph{Combust. Flame} 253: 112811 (2023).

\bibitem[Regele et~al.(2013)]{regele2013two}
Regele, J.D., Knudsen, E., Pitsch, H., Blanquart, G., \enquote{{A two-equation model for non-unity Lewis number differential diffusion in lean premixed laminar flames}}, \emph{Combust. Flame} 160(2): 240--250 (2013).

\bibitem[Berger et~al.(2022{\natexlab{c}})]{berger2022development}
Berger, L., Attili, A., Wang, J., Maeda, K., Pitsch, H., \enquote{Development of large-eddy simulation combustion models for thermodiffusive instabilities in turbulent hydrogen flames}, \emph{Proc. of the Summer Prog., Center for Turb. Research, Stanford University} p. 247 (2022{\natexlab{c}}).

\bibitem[Yao and Blanquart(2024)]{yao2024capturing}
Yao, M.X., Blanquart, G., \enquote{Capturing differential diffusion effects in large eddy simulation of turbulent premixed flames}, \emph{Proc. Combust. Inst.} 40(1-4): 105500 (2024).

\bibitem[Aspden(2017)]{aspden2017numerical}
Aspden, A.J., \enquote{A numerical study of diffusive effects in turbulent lean premixed hydrogen flames}, \emph{Proc. Combust. Inst.} 36(2): 1997--2004 (2017).

\bibitem[B{\"o}ttler et~al.(2022)]{bottler2022flamelet}
B{\"o}ttler, H., Chen, X., Xie, S., Scholtissek, A., Chen, Z., Hasse, C., \enquote{Flamelet modeling of forced ignition and flame propagation in hydrogen-air mixtures}, \emph{Combust. Flame} 243: 112125 (2022).

\bibitem[B{\"o}ttler et~al.(2023)]{bottler2023flamelet}
B{\"o}ttler, H., Lulic, H., Steinhausen, M., Wen, X., Hasse, C., Scholtissek, A., \enquote{Flamelet modeling of thermo-diffusively unstable hydrogen-air flames}, \emph{Proc. Combust. Inst.} 39(2): 1567--1576 (2023).

\bibitem[B{\"o}ttler et~al.(2024)]{bottler2024can}
B{\"o}ttler, H., Kaddar, D., Karpowski, T.J.P., Ferraro, F., Scholtissek, A., Nicolai, H., Hasse, C., \enquote{Can flamelet manifolds capture the interactions of thermo-diffusive instabilities and turbulence in lean hydrogen flames?—an a-priori analysis}, \emph{Int. J. Hydrogen Energy} 56: 1397--1407 (2024).

\bibitem[Berger et~al.(2022{\natexlab{d}})]{berger2022intrinsica}
Berger, L., Attili, A., Pitsch, H., \enquote{{Intrinsic instabilities in premixed hydrogen flames: Parametric variation of pressure, equivalence ratio, and temperature. Part 1--Dispersion relations in the linear regime}}, \emph{Combust. Flame} 240: 111935 (2022{\natexlab{d}}).

\bibitem[Rieth et~al.(2023)]{rieth2023effect}
Rieth, M., Gruber, A., Chen, J.H., \enquote{The effect of pressure on lean premixed hydrogen-air flames}, \emph{Combust. Flame} 250: 112514 (2023).

\bibitem[Esclapez et~al.(2023)]{PeleLMeX_JOSS}
Esclapez, L., Day, M., Bell, J., Felden, A., Gilet, C., Grout, R., {Henry de Frahan}, M., Motheau, E., Nonaka, A., Owen, L., Perry, B., Rood, J., Wimer, N., Zhang, W., \enquote{{PeleLMeX: an AMR Low Mach Number Reactive Flow Simulation Code without level sub-cycling}}, \emph{J. Open Source Softw.} 8(90): 5450 (2023).

\bibitem[Nonaka et~al.(2018)]{nonaka2018conservative}
Nonaka, A., Day, M.S., Bell, J.B., \enquote{{A conservative, thermodynamically consistent numerical approach for low Mach number combustion. Part I: Single-level integration}}, \emph{Combust. Theory Modell.} 22(1): 156--184 (2018).

\bibitem[Almgren et~al.(1998)]{almgren1998conservative}
Almgren, A.S., Bell, J.B., Colella, P., Howell, L.H., Welcome, M.L., \enquote{{A conservative adaptive projection method for the variable density incompressible Navier--Stokes equations}}, \emph{J. Comput. Phys.} 142(1): 1--46 (1998).

\bibitem[Day and Bell(2000)]{day2000numerical}
Day, M.S., Bell, J.B., \enquote{Numerical simulation of laminar reacting flows with complex chemistry}, \emph{Combust. Theory Modell.} 4(4): 535 (2000).

\bibitem[Burke et~al.(2012)]{burke2012comprehensive}
Burke, M.P., Chaos, M., Ju, Y., Dryer, F.L., Klippenstein, S.J., \enquote{{Comprehensive \ce{H2}/\ce{O2} kinetic model for high-pressure combustion}}, \emph{Int. J. Chem. Kinet.} 44(7): 444--474 (2012).

\bibitem[Berger et~al.(2019)]{berger2019characteristic}
Berger, L., Kleinheinz, K., Attili, A., Pitsch, H., \enquote{Characteristic patterns of thermodiffusively unstable premixed lean hydrogen flames}, \emph{Proc. Combust. Inst.} 37(2): 1879--1886 (2019).

\bibitem[Goodwin et~al.(2023)]{cantera}
Goodwin, D.G., Moffat, H.K., Schoegl, I., Speth, R.L., Weber, B.W., \enquote{Cantera: An object-oriented software toolkit for chemical kinetics, thermodynamics, and transport processes}, \url{https://www.cantera.org} (2023), version 3.0.0.

\bibitem[Law and Sung(2000)]{law2000structure}
Law, C.K., Sung, C.J., \enquote{Structure, aerodynamics, and geometry of premixed flamelets}, \emph{Prog. Energy Combust. Sci.} 26(4-6): 459--505 (2000).

\bibitem[Law(2006)]{law2006propagation}
Law, C.K., \enquote{Propagation, structure, and limit phenomena of laminar flames at elevated pressures}, \emph{Combust. Sci. Technol.} 178(1-3): 335--360 (2006).

\bibitem[Day et~al.(2009)]{day2009turbulence}
Day, M., Bell, J., Bremer, P.T., Pascucci, V., Beckner, V., Lijewski, M., \enquote{Turbulence effects on cellular burning structures in lean premixed hydrogen flames}, \emph{Combust. Flame} 156(5): 1035--1045 (2009).

\bibitem[Howarth et~al.(2024)]{howarth2024thermal}
Howarth, T.L., Day, M.S., Pitsch, H., Aspden, A.J., \enquote{Thermal diffusion, exhaust gas recirculation and blending effects on lean premixed hydrogen flames}, \emph{Proc. Combust. Inst.} 40(1-4): 105429 (2024).

\bibitem[Lehmann et~al.(2025)]{lehmann2025comprehensive}
Lehmann, T., Berger, L., Howarth, T.L., Gauding, M., Girhe, S., Dally, B.B., Pitsch, H., \enquote{Comprehensive linear stability analysis for intrinsic instabilities in premixed ammonia/hydrogen/air flames}, \emph{Combustion and Flame} 273: 113927 (2025).

\end{thebibliography}

\newpage
\section{Supplementary material}
\begin{figure}[h!]
    \centering
    \includegraphics[width=0.98\linewidth]{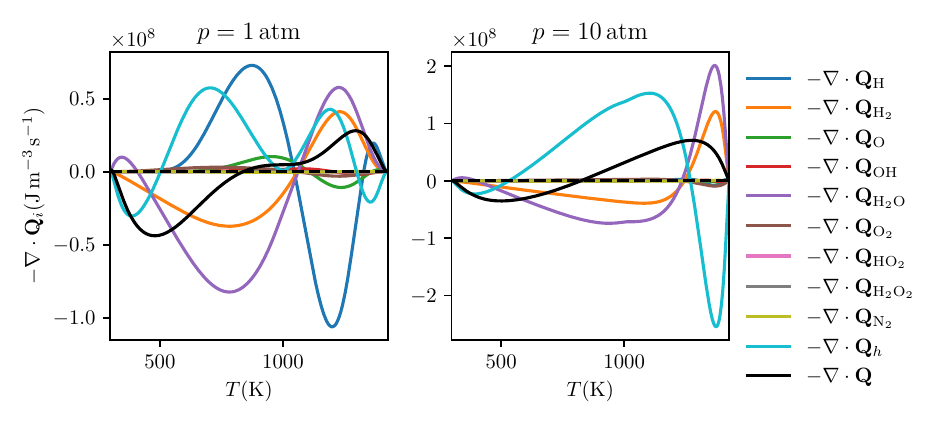}
    \caption{Enthalpy flux divergence terms $-\nabla \cdot \mathbf{Q}_{h}$ and $-\nabla \cdot \mathbf{Q}_{k}$ for all species $k$. The dashed line denotes 0.}
    \label{fig:enth_source_full}
\end{figure}

\end{document}